# Quality Assurance on Un-Doped CsI Crystals for the Mu2e Experiment

N. Atanov, V. Baranov, J. Budagov, Yu. I. Davydov, V. Glagolev, V. Tereshchenko, Z. Usubov
Joint Institute for Nuclear Research, Dubna, Russia

F. Cervelli, S. Di Falco, S. Donati, L. Morescalchi, E. Pedreschi, G. Pezzullo, F. Raffaelli, F. Spinella
INFN sezione di Pisa, Pisa, Italy

F. Colao, M. Cordelli, G. Corradi, E. Diociaiuti, R. Donghia, S. Giovannella, F. Happacher, M. Martini,
S. Miscetti, M. Ricci, A. Saputi, I. Sarra
Laboratori Nazionali di Frascati dell' INFN, Frascati, Italy

B. Echenard, D. G. Hitlin, C. Hu, T. Miyashita, F. Porter, L. Zhang, R.-Y. Zhu*
California Institute of Technology, Pasadena, California, USA

F. Grancagnolo, G. Tassielli
INFN sezione di Lecce, Lecce, Italy

P. Murat
Fermi National Accelerator Laboratory, Batavia, Illinois, USA

*Abstract*—**The Mu2e experiment is constructing a calorimeter consisting of 1,348 un-doped CsI crystals in two disks. Each crystal has a dimension of 34×34×200 mm$^3$, and is readout by a large area SiPM array. A series of technical specifications on mechanical and optical parameters was defined according to the calorimeter physics requirements. Pre-production CsI crystals were procured from three firms: Amcrys, Saint-Gobain and SIC. We report the quality assurance on crystal's scintillation properties and their radiation hardness against ionization dose and neutrons. With a fast decay time of about 30 ns and a light output of more than 100 p.e./MeV measured by a bi-alkali PMT, un-doped CsI crystals provide a cost-effective solution for Mu2e.**

## I. INTRODUCTION

AIMING at exploring lepton flavor violation, the Mu2e experiment [1] is constructing a calorimeter consisting of 1,348 un-doped cesium iodide (CsI) crystals of 34×34×200 mm$^3$ readout by a large area Silicon Photomultipliers (SiPM) array [2]. With a fast decay time of about 30 ns and a light output of more than 100 p.e./MeV measured by a bi-alkali PMT, un-doped CsI crystals provide a cost-effective solution for Mu2e. Technical specifications are defined for crystals quality assurance according to physics requirements:

- Crystal dimension tolerance: ±100 μm;
- Visual inspection: no cracks, chips, fingerprints, and free from inclusions and bubbles;
- Light output (LO) in 200 ns: > 100 p.e./MeV;
- FWHM Energy resolution for Na-22 peaks: < 45%;
- Light response uniformity (LRU): < 5%;
- Fast (200 ns)/Total (3,000 ns) Ratio: > 75%;
- Radiation Induced Noise (RIN) @1.8 rad/h: < 0.6 MeV;
- Normalized LO after 10/100 krad > 85%/60%.

These specifications define crystal's mechanical, optical and scintillation properties which would affect calorimeter performance. Crystal's longitudinal transmittance (LT) reveals optical absorption in the crystal bulk, so is a useful parameter for crystal quality control, particularly for investigations of radiation induced absorption. Its numerical value, however, is affected by crystal's surface quality and microscopic scattering centers in the crystal bulk, which do not affect calorimeter performance directly. Since CsI transmittance is affected by its hygroscopic surface quality, so the Mu2e specification does not include a transmittance requirement. For birefringent crystals, such as PWO, longitudinal transmittance is known to be different along different optical axis [3]. In this case, care should be taken to define different transmittance specification for crystals grown along different optical axis.

This work is supported by the U.S. Department of Energy, Office of High Energy Physics program under Award Number DE-SC0011925.

The corresponding author is with the California Institute of Technology, Pasadena, CA 91125 USA (e-mail: zhu@hep.caltech.edu).

In this paper, we report quality assurance on preproduction CsI crystals procured from three vendors: AMCRYS, Saint-Gobain (S-G) Corporation and Shanghai Institute of Ceramics (SIC). While scintillation properties, such as LO, FWHM energy resolution, LRU, F/T ratio and RIN, were measured for all CsI crystals, radiation hardness was measured for selected samples. The measured data are compared to the Mu2e technical specifications.

## II. Samples and Measurement Details

A total of 72 preproduction CsI crystals of $34\times34\times200$ mm$^3$ were procured from three vendors: AMCRYS, S-G and SIC with 24 crystals from each vendor. They were characterized at Caltech and LNF with 36 in each lab. The results from two labs are consistent. Fig. 1 is a photo showing 36 preproduction crystals, arranged in an order of Amcrys, S-G and SIC from the left to the right.

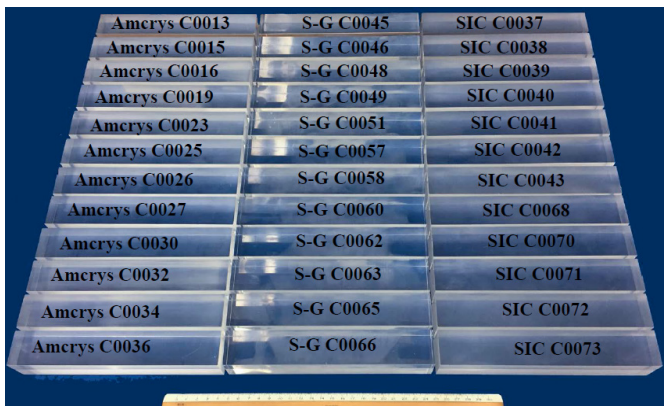

Fig. 1. 36 CsI crystals of $34\times34\times200$ mm$^3$ from Amcrys, S-G and SIC.

Crystals were wrapped with two layers of Tyvek paper of 150 µm with a selected end coupled to a bi-alkali PMT Hamamatsu R2059 via an air gap. The coupling end was chosen to provide a better LRU. Pulse height spectra were measured by using 0.511 MeV γ-rays from a $^{22}$Na source with a systematic uncertainty for the peak determination of about 1%. The LO and FWHM resolution are defined as the average of seven points measured along the crystal length with 200 ns integration time.

The LRU is defined as the standard deviation (rms) of the seven points. The LO was also measured as a function of the integration time at the point of 2.5 cm from the PMT, from which the F/T ratio is determined [4].

The radiation induced photocurrent was measured as the anode current during irradiation at a dose rate of 2 rad/h, and was used to extract the crystal's RIN at 1.8 rad/h. Radiation damage in both transmittance and LO was measured for two CsI crystals randomly selected from each vendor after 10 and 100 krad. In the photocurrent and LO measurements, crystals were with the same wrapping and air coupled to the same Hamamatsu R2059 PMT.

Longitudinal transmittance was measured by a PerkinElmer Lambda 950 spectrophotometer with 0.15% precision for selected crystals before and after irradiation.

## III. Basic Properties and their Correlations

Figs. 2 shows a summary of the LO in 200 ns for all 72 preproduction crystals together with the Mu2e specification of 100 p.e./MeV (red dashed lines). All crystals satisfy this specification. Crystals from SIC have the highest LO, while crystals from S-G are featured with the best overall consistency.

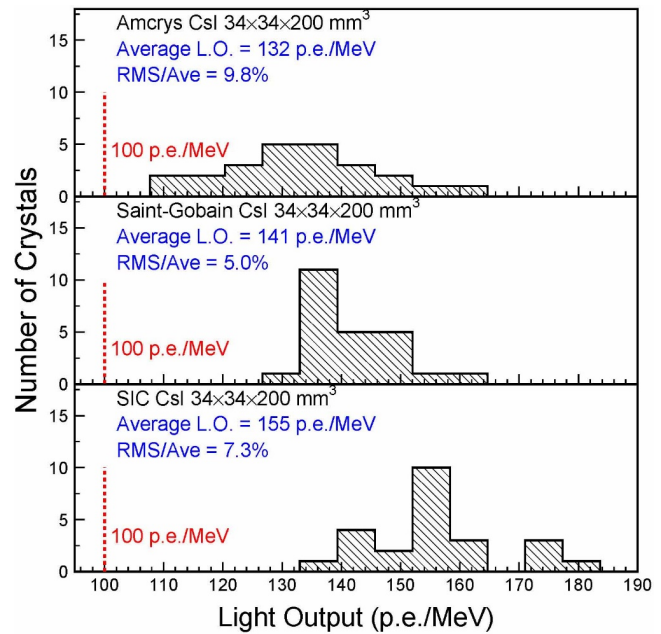

Fig. 2. A summary of the LO in 200 ns for 72 preproduction crystals compared to the Mu2e specification of 100 p.e./MeV (red dashed lines).

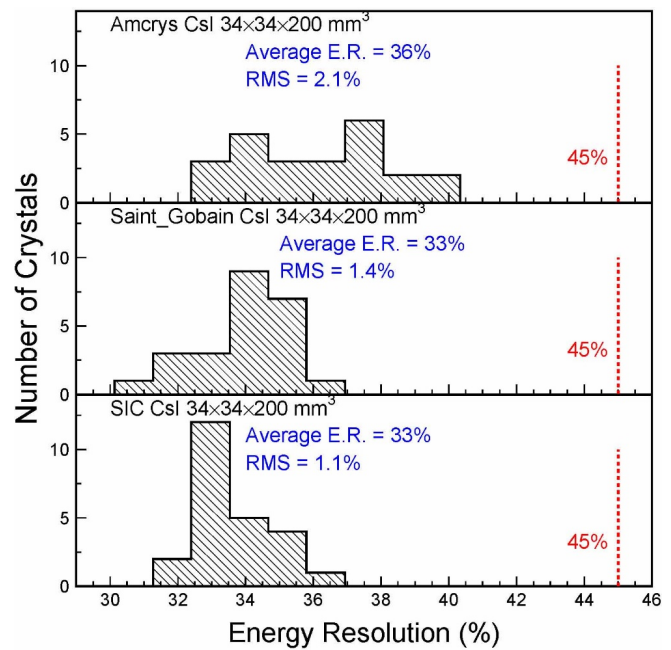

Fig. 3. A summary of the FWHM energy resolution for 72 preproduction crystals compared to the Mu2e specification of 45% (red dashed lines).

Fig. 3 shows a summary of the FWHM energy resolution for all 72 preproduction crystals together with the Mu2e specification of 45% (red dashed lines). All crystals satisfy this Mu2e specification. Crystals from SIC have the best energy resolution. This is consistent with the LO result.

Fig. 4 shows a summary of the LRU for all 72 preproduction crystals together with the Mu2e specifications of 5% (red dashed lines). Most crystals satisfy this specification, except two crystals from SIC. Crystals from Amcrys have the best LRU and overall consistency.

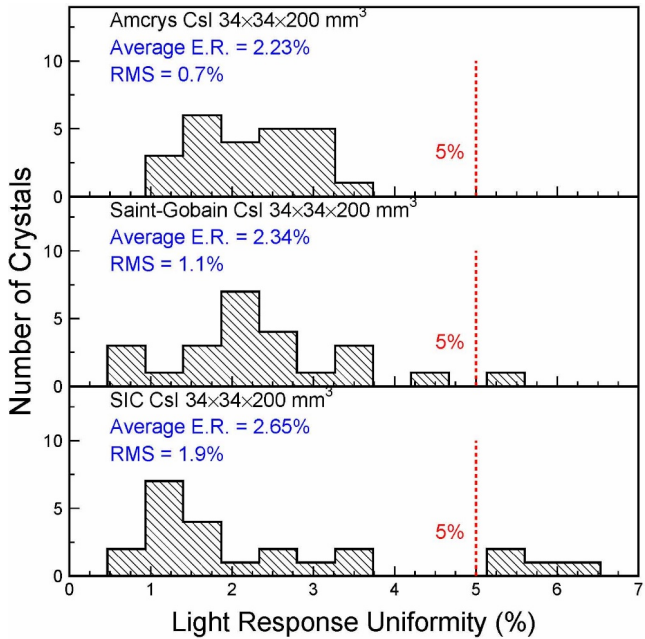

Fig. 4. A summary of the LRU for 72 preproduction crystals compared to the specification of 5% (red dashed lines).

Fig. 5 shows a summary of the F/T ratio for all 72 preproduction crystals together with the Mu2e specification of 75% (red dashed lines). Crystals from S-G show the best F/T and excellent consistency. Over half of the crystals from Amcrys fail this specification.

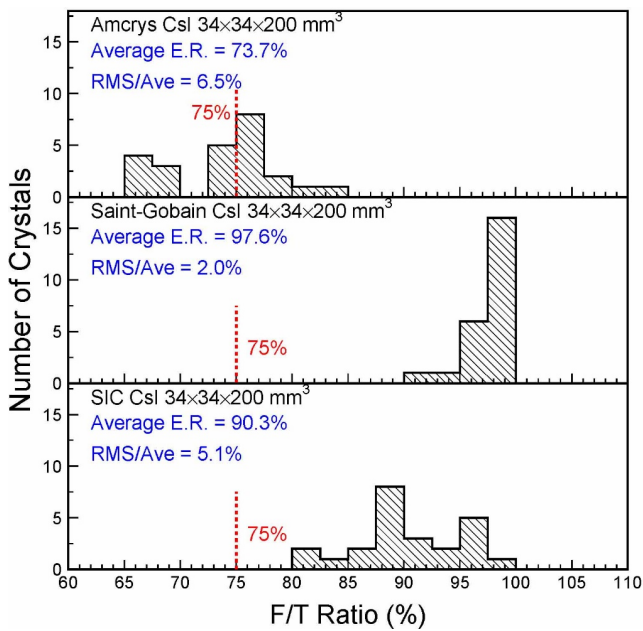

Fig. 5. A summary of the F/T ratio for 72 preproduction crystals compared to the specification of 75% (red dashed lines).

Figs. 6 and 7 show the correlations between the FWHM energy resolution versus the LO and the F/T ratio respectively for all 72 preproduction crystals. These good correlations indicate the importance of controlling the slow scintillation component for undoped CsI crystals [4]. The FWHM energy resolution seems to have a lower limit at 32%, which is believed to be intrinsic for mass-produced undoped CsI crystals of the Mu2e size.

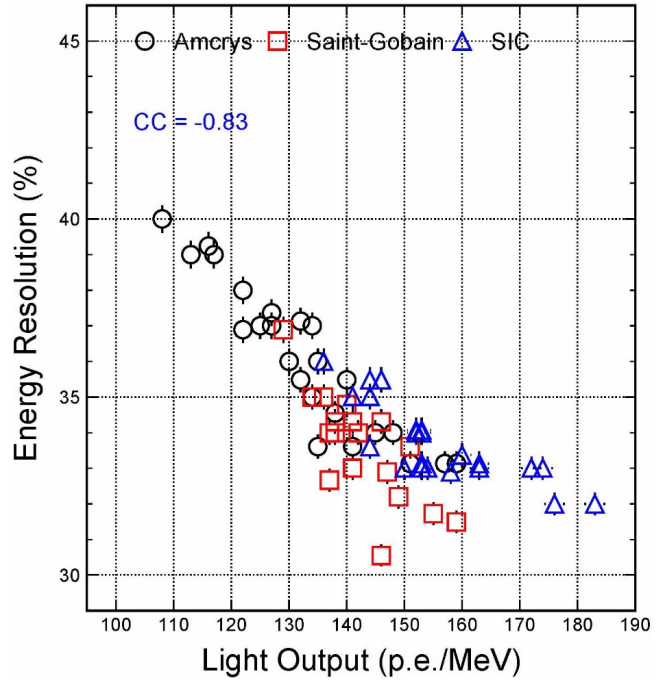

Fig. 6. Correlations between the LO and the FWHM energy resolution for 72 crystals, where CC is the correlation coefficient.

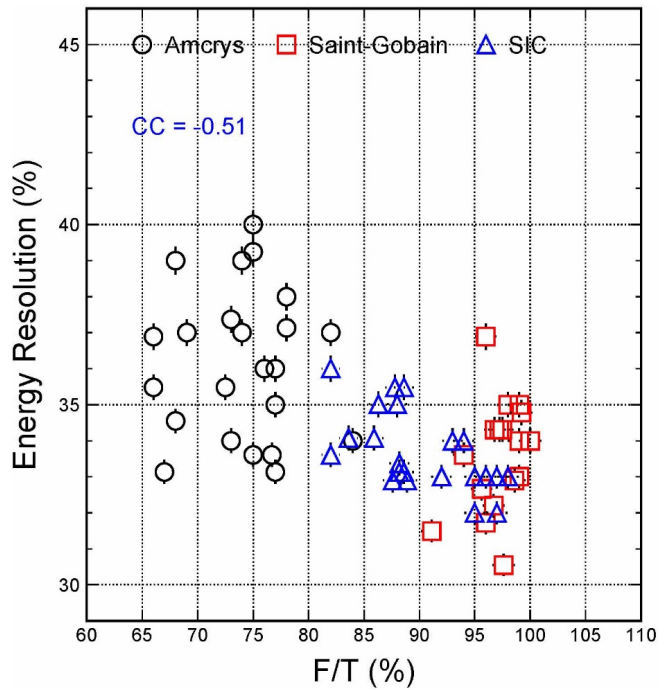

Fig. 7. Correlations between the F/T ratio and the FWHM energy resolution for 72 crystals, where CC is the correlation coefficient.

## IV. Gamma-ray Induced Photocurrent and Readout Noise

Fig. 8 shows a setup used to measure Co-60 γ-ray induced photocurrent for crystals under irradiation at a dose rate of 2 rad/h, which is compatible with the 1.8 rad/h expected in the hottest region of the Mu2e calorimeter [2].

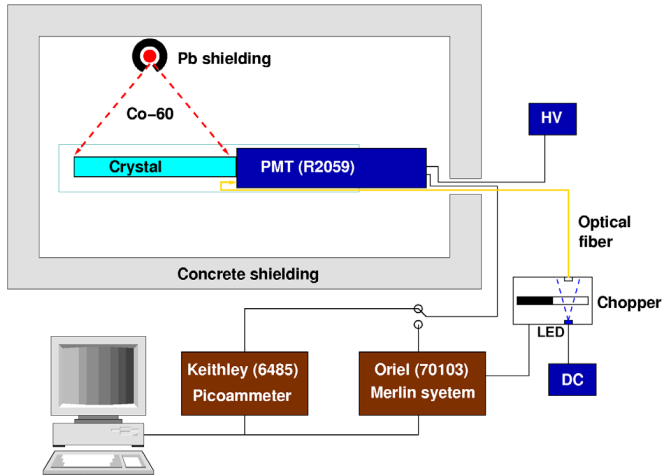

Fig. 8. A schematic showing the setup used to measure the γ-ray induced photocurrent.

The RIN is derived as the standard deviation of photoelectron numbers ($Q$) in the readout gate normalized to the LO [4]:

$$RIN = \frac{\sqrt{Q}}{LO} \quad (MeV). \quad (1)$$

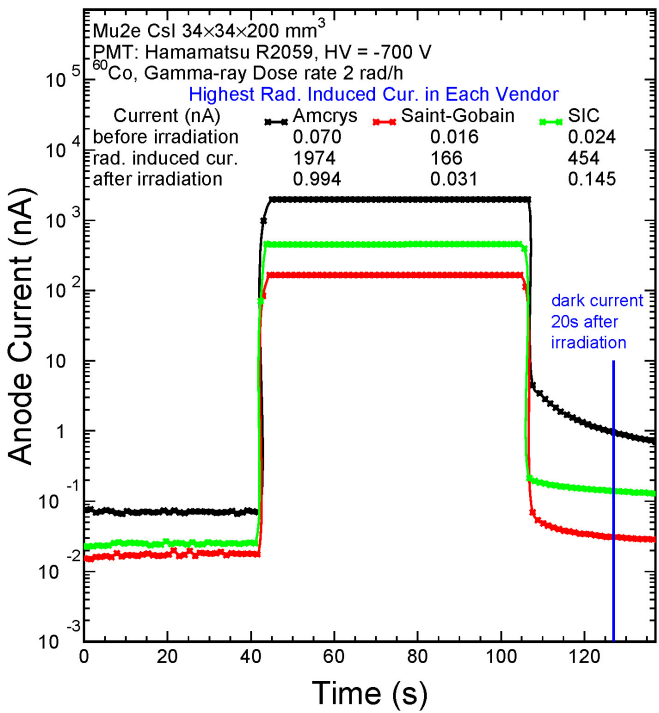

Fig. 9. Histories of the highest photocurrent from 3 vendors.

Fig. 9 shows histories of the photocurrent measured for 3 CsI crystals with the highest γ-ray induced photocurrent from each vendor. The dark current, the γ-ray induced photocurrent and the afterglow were measured before, during and after irradiation respectively for about 40, 65 and 35 s. Crystals from AMCRYS show the highest dark current, γ-ray induced current and afterglow. S-G CsI crystals show the lowest γ-ray induced photocurrent, about one order of the magnitude smaller than that of AMCRYS.

Fig. 10 shows a summary of the RIN values measured for all 72 preproduction crystals together with the specification (red dashed lines) of 0.6 MeV. S-G crystals have the lowest γ-ray induced noise and the best consistency, while AMCRYS crystals show the highest γ-ray induced noise. 14 out of the all 24 AMCRYS crystals fail the RIN specification.

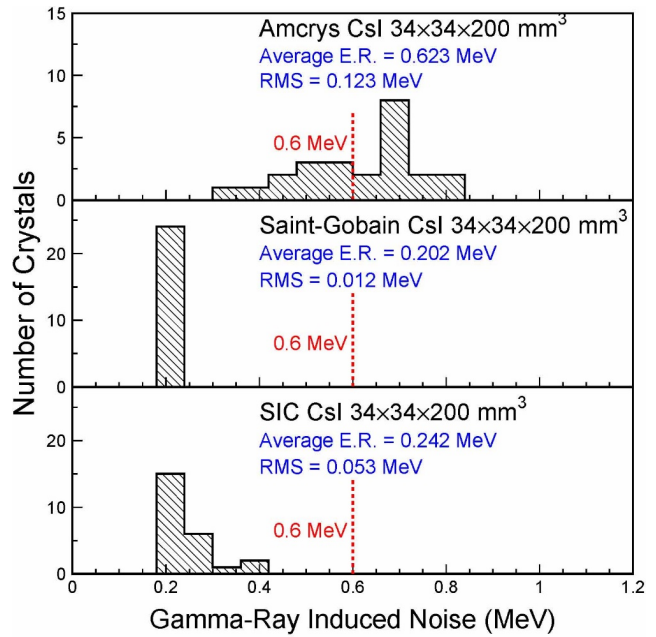

Fig. 10. A summary of the γ-ray induced readout noise for 72 preproduction crystals together with the Mu2e specification of 0.6 MeV (red dashed lines).

Fig. 11 shows the correlation between the γ-ray induced photocurrent and the dark current. A perfect correlation with correlation coefficient of 99% hints the same origin of these two currents.

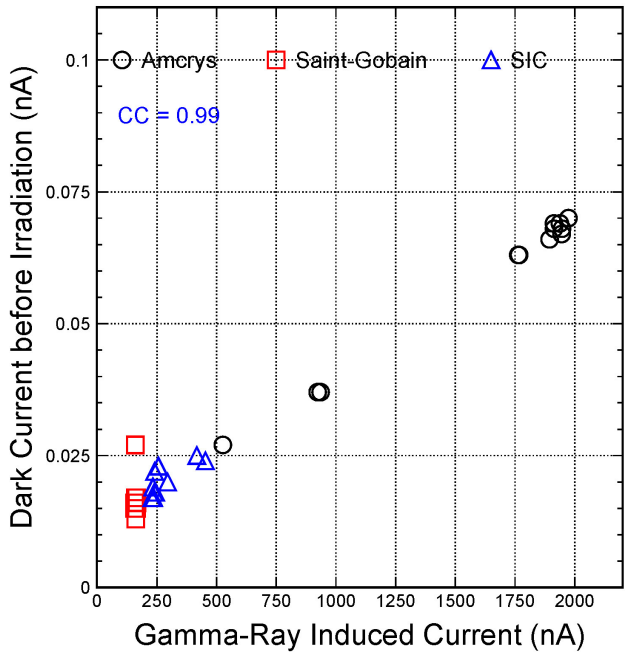

Fig. 11. Correlation between the γ-ray induced current and the dark current, where CC refers to the correlation coefficient.

Figs. 12 and 13 show correlations between the γ-ray induced noise and the dark current versus the F/T ratio. Excellent correlations are confirmed. This is in an addition to the correlation between the F/T ratio versus light output and resolution, enhancing the importance to reducing or eliminating the slow component. This result is consistent with our previous study on un-doped CsI crystal samples procured before the preproduction [4].

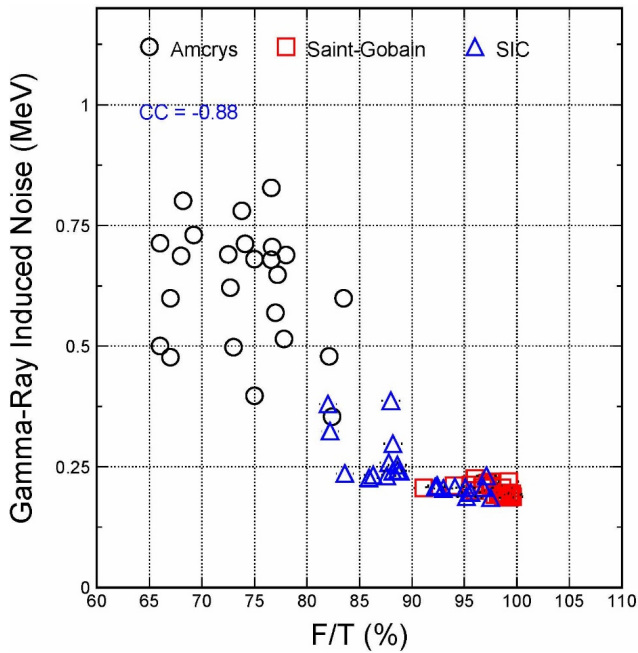

Fig. 12. Correlation between the F/T ratio and γ-ray induced noise, where CC is the correlation coefficient.

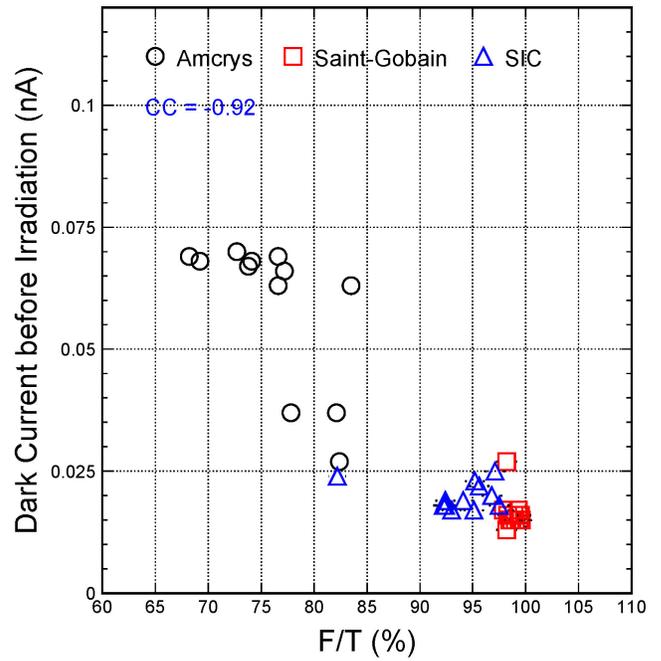

Fig. 13. Correlation between the F/T ratio and dark current, where CC is the correlation coefficient.

## V. GAMMA-RAY INDUCED RADIATION DAMAGE IN SIX PREPRODUCTION CsI CRYSTALS

Fig. 14 shows the LO of six preproduction CsI crystals after 10 and 100 krad. Most of these crystals have LO more than 100 p.e./MeV after 100 krad irradiation, promising a robust CsI calorimeter for Mu2e.

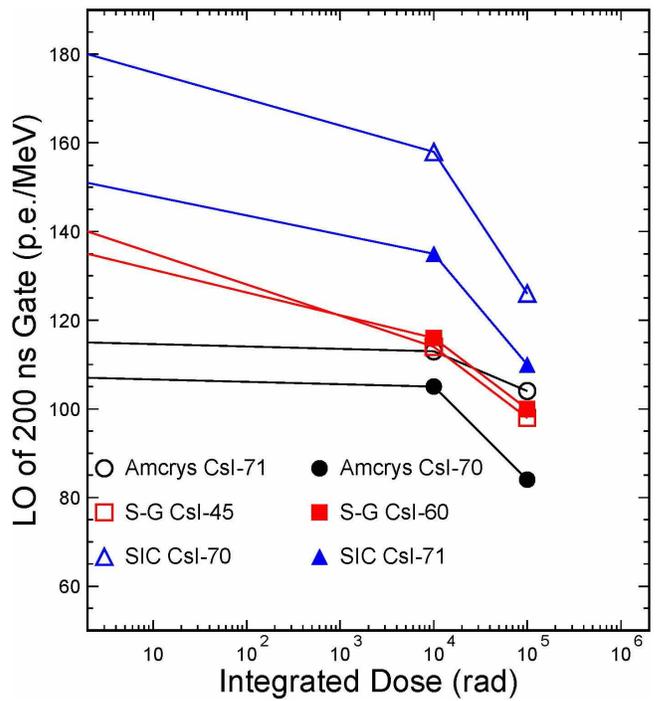

Fig. 14. LO after γ-ray irradiation are shown for six CsI crystals.

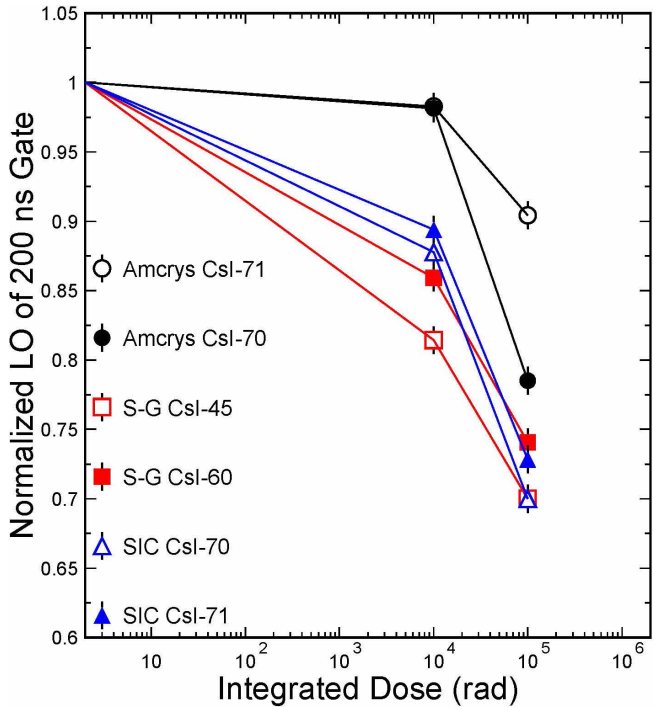

Fig. 15. Normalized LO loss after γ-ray irradiation is shown for six crystals.

Fig. 15 shows the normalized LO for six crystals after 10 and 100 krad. All crystals meet the Mu2e radiation damage specifications, except one Saint-Gobain sample (#45) which does not meet damage specification after 10 krad but meets that after 100 krad.

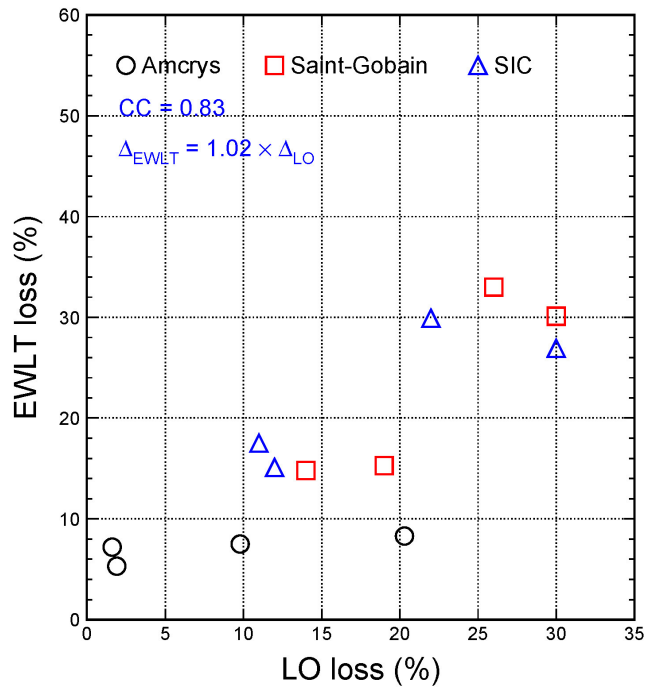

Fig. 16. Correlation between the losses of the LO and the EWLT is shown for 6 CsI crystals.

Fig. 16 shows the correlation between the LO loss versus the loss of the emission weighted longitudinal transmittance (EWLT), which is defined as:

$$EWLT = \int LT(\lambda) Em(\lambda) d\lambda \, / \int Em(\lambda) d\lambda. \quad (2)$$

The EWLT value provides a numerical representation of the LT data across the emission spectrum. The good correlation observed between the LO losses versus the transmittance losses indicates that the LO variation of undoped CsI crystals can be corrected by measuring crystal's transparency with a light monitoring system.

## VI. SUMMARY

72 preproduction CsI crystals from AMCRYS, S-G and SIC are characterized at Caltech and LNF, and are compared to the Mu2e technical specifications. AMCRYS crystals have the best uniformity, but the poorest light output, FWHM energy resolution and F/T ratio. About half AMCRYS crystals do not meet the F/T and RIN specifications. Saint-Gobain crystals have the best F/T ratio and overall consistency. One Saint-Gobain crystal does not meet radiation damage spec after 10 krad but meets that after 100 krad. SIC crystals have the best light output and energy resolution, but the poorest uniformity. Two SIC crystals do not meet the uniformity specification.

Correlations are observed between the LO, the FWHM energy resolution and the F/T ratio, indicating the importance of slow component control, which is believed to be raw material purity and defects related [4]. Correlations are also observed between the dark current, the radiation induced current/noise and the F/T ratio, enhancing the need to control the F/T ratio.

Most crystals have LO larger than 100 p.e./MeV after 100 krad, promising a robust CsI calorimeter for the Mu2e experiment at Fermilab. Correlation are also observed between the variations of the EWLT and the LO, indicating that a light monitoring system is useful for the Mu2e CsI calorimeter to correct variations of the LO by measuring variations of crystal's transparency.

Based upon this investigation S-G and SIC are selected to be the suppliers by the Mu2e experiment to construct the Mu2e CsI calorimeter. We will keep a close communication with the suppliers for crystals quality assurance during the construction.

## ACKNOWLEDGMENT

We are grateful for the vital contributions of the Fermilab staff and the technical staff of the participating institutions. This work was supported by the US Department of Energy; the Italian Istituto Nazionale di Fisica Nucleare; the US National Science Foundation; the Ministry of Education and Science of the Russian Federation; the Thousand Talents Plan of China; the Helmholtz Association of Germany; and the EU Horizon 2020 Research and Innovation Program under the Marie Sklodowska-Curie Grant Agreement N.690385. Fermilab is operated by Fermi Research Alliance, LLC under Contract No. De-AC02-07CH11359 with the US Department of Energy, Office of Science, Office of High Energy Physics.